\documentclass[conference]{IEEEtran}
\IEEEoverridecommandlockouts
\usepackage{cite}
\usepackage{amsmath,amssymb,amsfonts}
\usepackage{graphicx}
\usepackage{textcomp}
\usepackage{xcolor}
\usepackage{booktabs}
\usepackage{multirow}
\usepackage{listings}
\usepackage[hyphens]{url}
\usepackage{algorithm}
\usepackage{algpseudocode}
\usepackage{colortbl}
\usepackage{pifont}
\usepackage{balance}

\definecolor{cpass}{RGB}{213,245,227} 
\definecolor{ccomp}{RGB}{250,219,219} 
\definecolor{cfunc}{RGB}{253,235,208} 
\definecolor{ctime}{RGB}{235,237,239}

\def\BibTeX{{\rm B\kern-.05em{\sc i\kern-.025em b}\kern-.08em
    T\kern-.1667em\lower.7ex\hbox{E}\kern-.125emX}}

\begin{document}

\title{RefEvo: Agentic Design with Co-Evolutionary Verification for Agile Reference Model Generation}

\author{%
  \parbox{\textwidth}{\centering
    Yifan Zhang\textsuperscript{1,2},
    Jianmin Ye\textsuperscript{1,2},
    Jiahao Yang\textsuperscript{2},
    Xi Wang\textsuperscript{1,2*}
    \vspace{10pt} \\ 
    \textsuperscript{1}School of Integrated Circuits, Southeast University, China \\
    \textsuperscript{2}National Center of Technology Innovation for EDA, China \\
    Email: \{yifan.zhang, jianmin\_y, xi.wang\}@seu.edu.cn, yangjiahao@nctieda.com\\
    \textsuperscript{*}Corresponding author.
  }
}

\maketitle

\begin{abstract}
As the complexity of Systems-on-Chip (SoC) escalates, the ``shift-left'' strategy necessitates the rapid development of high-fidelity reference models (e.g., in SystemC) for early architecture exploration and verification. While Large Language Models show promise in code generation, applying them to hardware modeling faces distinct challenges: (1) rigid, static workflows fail to adapt to varying design complexities, causing inefficiency; (2) context overflow in long-turn interactions leads to the catastrophic forgetting of critical specifications; and (3) the ``Coupled Validation Failure'' problem—where generated TBs falsely verify flawed models due to shared hallucinations—severely undermines reliability. To address these limitations, we introduce \textbf{RefEvo}, a dynamic multi-agent framework designed for agile and reliable modeling. RefEvo features three key innovations: (1) A \textbf{Dynamic Design Planner} that autonomously decomposes specifications and constructs tailored execution workflows based on semantic complexity; (2) A \textbf{Co-Evolutionary Verification Mechanism}, which employs a Dialectical Arbiter to simultaneously rectify the model and verification logic against the specification (Spec) oracle, effectively mitigating false positives; and (3) A \textbf{Spec Anchoring Strategy} for lossless context compression. Evaluated on a diverse benchmark of 20 hardware modules, RefEvo achieves a \textbf{95\%} pass rate, outperforming static baselines by a large margin. Furthermore, our context optimization reduces token consumption by an average of \textbf{71.04\%}, achieving absolute savings of over \textbf{70,000 tokens} per session for complex designs while maintaining 100\% specification recall.
\end{abstract}


\section{Introduction}
In the modern Electronic Design Automation (EDA) landscape, high-level reference models written in languages like SystemC or C++ have become critical assets \cite{2006Design}. These models serve as the ``golden standard'' for verifying Register Transfer Level (RTL) implementations and act as virtual prototypes for early software development \cite{hardware_verification_llm2025, chatmodel2025,2018Different}. Their importance is further amplified by the industry-wide adoption of the ``shift-left'' methodology, which mandates early validation through virtual platforms to enable hardware-software co-verification prior to silicon fabrication. However, despite their necessity, manually crafting these high-fidelity models remains a labor-intensive and error-prone process, creating a significant bottleneck in the agile hardware development cycle \cite{2015Trends,2015Case}.

To alleviate this engineering burden, recent advancements in Large Language Models (LLMs) have sparked intense interest in automating EDA tasks \cite{chipnemo2023, llm_verification_survey2026}. While LLMs like GPT-4 have demonstrated remarkable proficiency in generating generic software, applying them to the strict constraints of hardware modeling presents unique challenges. Within the hardware domain, the majority of research progress has been concentrated on RTL code generation, exemplified by approaches such as RTLCoder \cite{rtlcoder2024}, ChatChisel \cite{10618053}, ReChisel \cite{niu2025rechisel}, and ChatCPU \cite{10.1145/3649329.3658493}. Complementing these generation efforts, benchmarking studies \cite{verilog_benchmark2022} have also established robust evaluation frameworks specifically for assessing LLM capabilities in Verilog.

\begin{figure}[t]
\centerline{\includegraphics[width=1\linewidth]{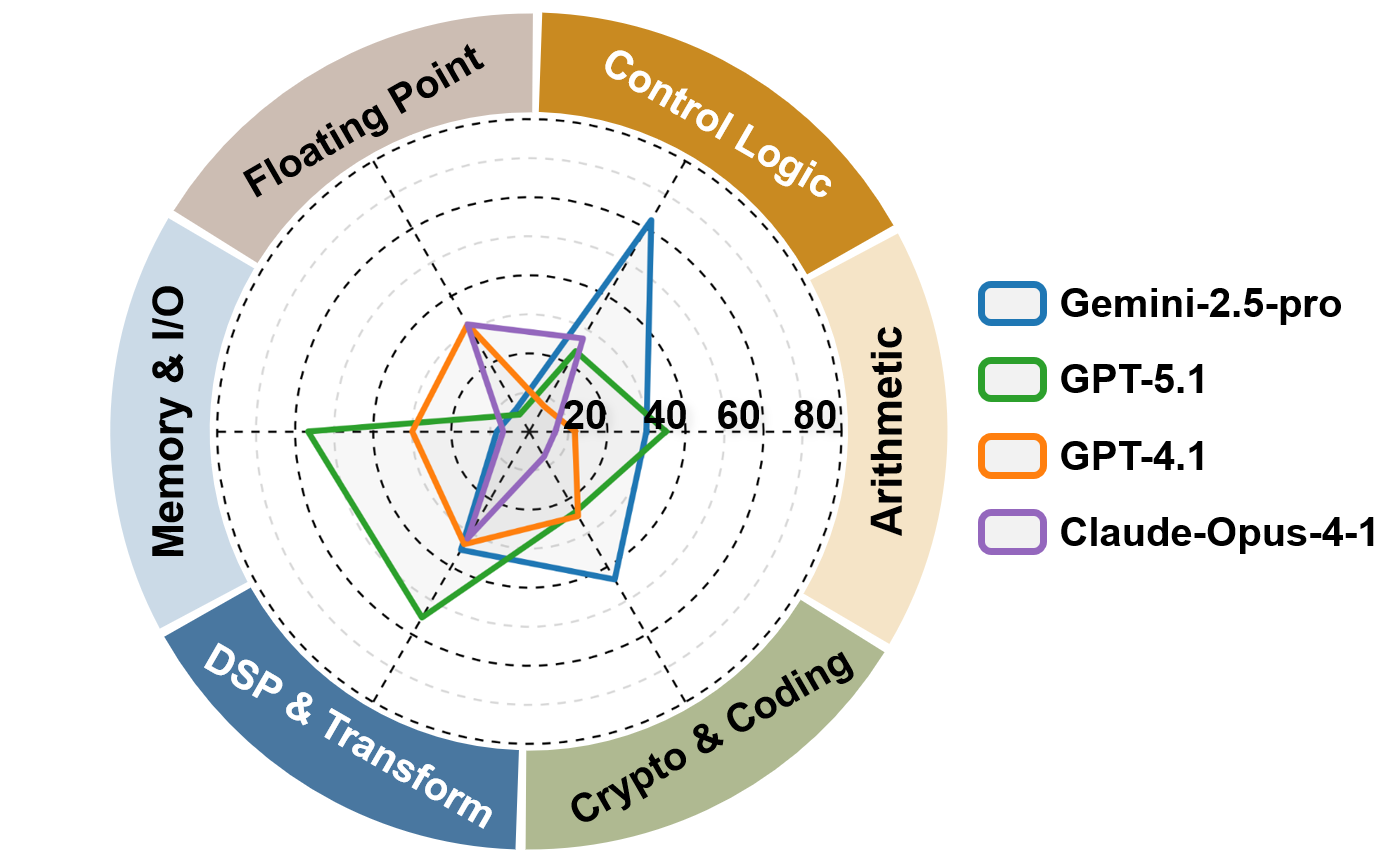}} 
\caption{Performance capability of state-of-the-art LLMs using optimized prompt engineering and structured generation workflows (without Co-Evolutionary Verification). The radar chart highlights the significant struggle of current models in generating correct SystemC reference models across various hardware domains, particularly in Control Logic and Memory \& I/O.}
\label{fig:intro_bg}
\end{figure}

However, SystemC modeling, which requires precise handling of transaction-level modeling (TLM) semantics, bit-widths, concurrency, and delta-cycle timing, remains a distinct challenge \cite{2016Towards}. As illustrated in Fig.~\ref{fig:intro_bg}, even when equipped with specialized design workflows and sophisticated prompt engineering\cite{2023Large,2023The}, advanced models struggle significantly to produce functional SystemC models across various hardware domains. A naive or purely structured application of LLMs often results in code that looks plausible but fails to compile or simulate correctly. We identify three primary hurdles hindering the practical adoption of LLMs in this domain:

\textbf{Syntactic and Semantic Correctness:} Hardware models must adhere to strict event-driven simulation semantics. LLMs frequently generate code with subtle API misuses or missing library dependencies, leading to compilation failures that naive generation cannot resolve.

\textbf{Coupled Validation Failure:} A critical risk arises when LLMs generate both the design under test (DUT) and the testbench (TB). Traditional static flows \cite{2020A} often lead to a sycophantic scenario where the TB is inadvertently tailored to match the DUT's hallucinations, resulting in a false-positive ``PASS''. While introducing manual human-written TBs could mitigate this, it re-introduces the very bottleneck automation seeks to eliminate, as addressed by recent verification frameworks like ChatTest \cite{wan2026chattest} and FIXME \cite{FIXME}. Furthermore, in early agile stages where specifications change frequently, manual TBs suffer from \textit{error propagation}, where outdated verification logic fails to catch new design errors \cite{2024UVLLM,2021Effective}. Therefore, an automated mechanism to validate the validator itself is essential.

\textbf{Context Limitations:} As design complexity grows, the extensive interaction logs required for iterative refinement often exceed the context window of LLMs \cite{2023Self,chipmind,Fu2026ChatSVABS}. This causes the model to suffer from \textit{Catastrophic Forgetting}, losing track of initial specifications such as register maps or interface protocols during long interaction sessions.

To overcome these challenges, we propose \textbf{RefEvo}, an agent-based framework designed for the agile generation of reliable hardware reference models. Unlike static generation approaches, RefEvo treats generation as a dynamic, co-evolutionary process. Building on the collaborative paradigm of modern multi-agent frameworks \cite{agentmesh2025, multiagent_eda2025, multiagent_complex2025}---where planners, coders, and debuggers work in concert---RefEvo introduces domain-specific innovations to address the aforementioned problems:

\begin{itemize}
    \item \textbf{Dynamic Design Planner:} To address workflow rigidity, we introduce a planner that acts as a ``Brain'', analyzing semantic complexity to autonomously decompose specifications and construct bespoke execution workflows.
    \item \textbf{Co-Evolutionary Verification:} To mitigate coupled validation failures, we propose a dialectical mechanism where a dedicated Arbiter Agent validates both the code and the TB against the specification oracle, significantly reducing false positives.
    \item \textbf{Spec Anchoring Context Management:} To solve catastrophic forgetting, we implement a strategy that anchors the immutable specification while compressing interaction history, ensuring 100\% recall of critical constraints.
\end{itemize}

The remainder of this paper is organized as follows: Section~\ref{sec:two} reviews related work. Section~\ref{sec:three} details the RefEvo methodology. Section~\ref{sec:four} presents the experimental results and analysis. Section~\ref{sec:five} concludes the paper.

\section{Related Work}
\label{sec:two}

\begin{figure*}[htbp]
\centerline{\includegraphics[width=1\linewidth]{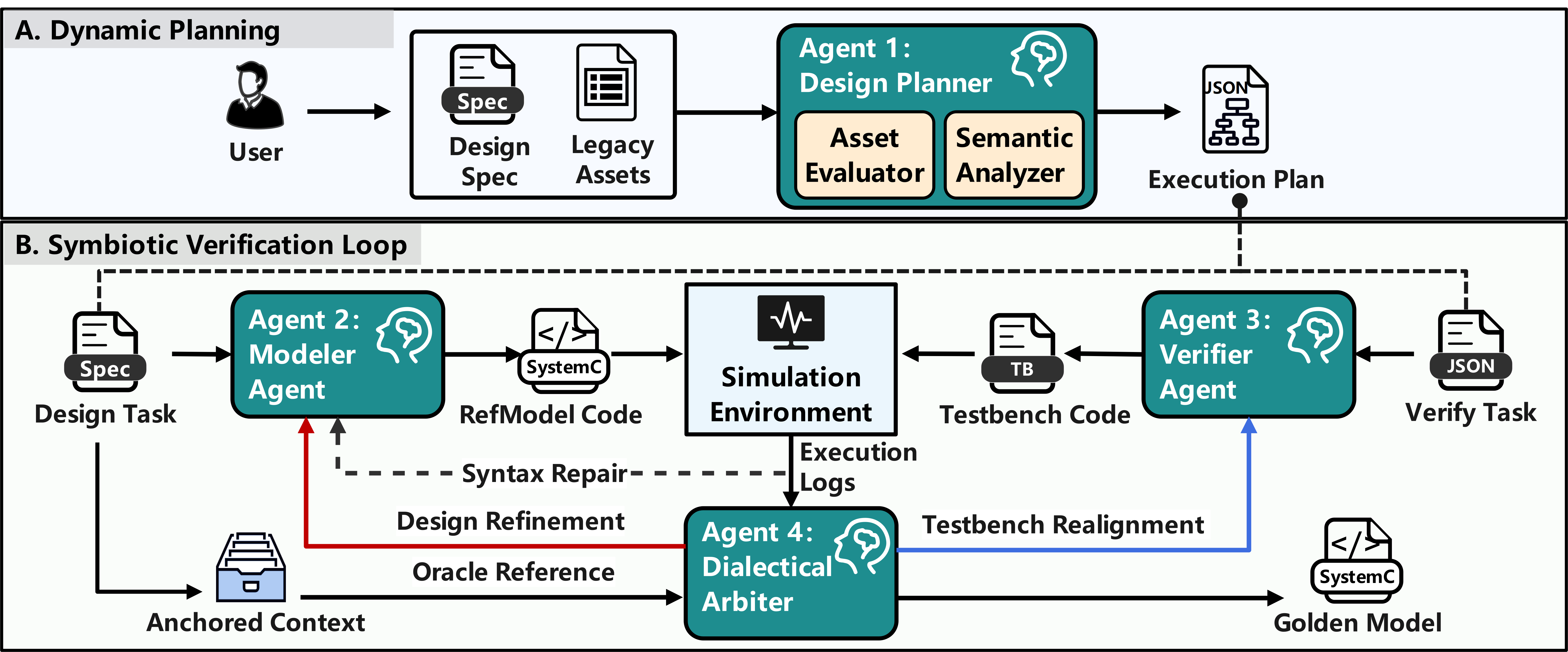}} 
\caption{The logical architecture of RefEvo. (A) The Dynamic Planning phase where Agent 1 analyzes complexity to construct an execution plan. (B) The Symbiotic Verification Loop where the Modeler (Agent 2) and Verifier (Agent 3) co-evolve under the supervision of the Dialectical Arbiter (Agent 4).}
\label{fig:fig1_arch}
\end{figure*}

\subsection{LLM-Based Hardware Code Generation}
The application of LLMs in hardware design has gained significant traction. Early benchmarking efforts \cite{verilog_benchmark2022} established evaluation frameworks demonstrating that LLMs could generate syntactically correct Verilog code. Tools like RTLCoder \cite{rtlcoder2024} have demonstrated that fine-tuned LLMs can generate syntactically correct RTL code, while iDSE \cite{Li2025iDSEND} also explored the integration of LLMs in high-level synthesis design space exploration. However, these works primarily focus on RTL generation. High-level modeling (SystemC/C++), which requires a higher level of abstraction and distinct simulation semantics, remains under-explored. RefEvo bridges this gap by targeting the specific challenges of reference model generation.

A critical challenge in LLM-based code generation is the hallucination problem. Recent work on code hallucination detection has categorized errors into mapping, naming, resource, and logic errors. Our work addresses the more subtle ``Coupled Validation Failure'', where both the model and its TB may be flawed. Unlike prior works that rely on fixed TBs, RefEvo introduces a co-evolutionary approach to dynamically refine the verification logic itself.

\subsection{Automated Debugging and Program Repair}
Automated program repair (APR) has emerged as a critical research area \cite{apr_survey2024}. Recent work has demonstrated the effectiveness of conversation-based APR \cite{conversation_apr2024}, where iterative dialogue with LLMs can fix bugs at remarkably low cost. Practical APR approaches \cite{practical_apr2024} have focused on reducing reliance on extensive test suites. In the hardware domain, recent works have attempted to use LLMs to fix RTL bugs based on compiler feedback. However, most existing approaches assume a fixed, correct TB. RefEvo distinguishes itself by introducing a dialectical arbiter that validates the TB against the specification, ensuring that "repairs" do not simply mask bugs.

\subsection{Dynamic Planning and Agent Reasoning}
The ability to dynamically decompose complex tasks is fundamental to effective agentic systems. TDAG \cite{tdag2025} introduced a framework that dynamically decomposes tasks into subtasks. Recent advances \cite{advancing_agentic2024} have integrated dynamic task decomposition with tool selection. RefEvo's Dynamic Design Planner draws inspiration from these advances, employing semantic complexity analysis to construct tailored execution workflows.

\section{Methodology}
\label{sec:three}

RefEvo is structured as a hierarchical multi-agent framework comprising three functional components: \textbf{Dynamic Planning}, the \textbf{Symbiotic Verification Loop}, and \textbf{Spec-Anchored Context Management}. The overall execution flow coordinates four specialized agents to transform ambiguous specifications into verified SystemC models.

\subsection{Dynamic Task Planning}
Traditional LLM-based EDA flows often enforce a rigid, linear execution sequence, leading to token wastage for simple logic and verification failures for complex systems. As illustrated in Fig.~\ref{fig:fig1_arch}A, the \textbf{Design Planner (Agent 1)} serves as the methodological orchestrator of the system.

Upon receiving the \textit{Design Specification} and optional \textit{Legacy Assets}, Agent 1 performs a dual-stage assessment:
\begin{itemize}
    \item \textbf{Asset Evaluator:} It examines existing codebases or Verification Intellectual Property to determine reusability, minimizing redundant generation.
    \item \textbf{Semantic Analyzer:} It evaluates the design's complexity across three dimensions: \textit{Interface Protocol} (e.g., AXI vs. GPIO), \textit{State Space} (e.g., complex FSMs), and \textit{Concurrency} (e.g., multi-clock domains).
\end{itemize}

The output of this component is a structured \textbf{Execution Plan} (JSON), which defines the task decomposition and selects the appropriate routing strategy.

\subsection{Symbiotic Verification Loop}
Guided by the strategy defined in the Execution Plan, the framework enters the execution phase. The core execution engine, shown in Fig.~\ref{fig:fig1_arch}B, employs a symbiotic relationship between generation and verification. This component addresses the \textit{Coupled Validation Failure} by decoupling the modeling and verification responsibilities.

\subsubsection{Dual-Generator Setup}
The framework dispatches two concurrent tasks: the \textbf{Modeler Agent (Agent 2)} generates the SystemC Reference Model, while the \textbf{Verifier Agent (Agent 3)} constructs the TB. These components are integrated into a \textbf{Simulation Environment} for compilation and execution.

\subsubsection{The Dialectical Mechanism}
The \textbf{Dialectical Arbiter (Agent 4)} acts as the central authority. By prompting the Arbiter to strictly verify the specific constraints anchored in the Spec against the simulation trace, it analyzes the \textit{Execution Logs} against the \textit{Oracle Reference} (the anchored specification) to resolve mismatches through four distinct routing paths:
\begin{itemize}
    \item \textbf{Syntax Repair:} For compilation errors, the Arbiter extracts error signatures and triggers an implicit repair loop for Agent 2 or 3.
    \item \textbf{Design Refinement (Red Path):} If a functional mismatch violates the specification, the Arbiter identifies the flaw in the DUT and directs Agent 2 to rectify the model logic.
    \item \textbf{TB Realignment (Blue Path):} If the DUT aligns with the specification but the TB asserts a false failure, the Arbiter identifies the hallucination in the verification logic and directs Agent 3 to correct the TB. This path is critical for breaking the symmetry of shared hallucinations.
    \item \textbf{Success (Green Path):} Once the Arbiter confirms functional equivalence, it outputs the \textbf{Golden Model} (verified SystemC and TB).
\end{itemize}

\subsection{Spec Anchoring Context Management}
To support the long-turn reasoning required in verification loop, we implement an Optimization component to prevent \textit{Catastrophic Forgetting}. As illustrated in Fig.~\ref{fig:spec_anchor}, we partition the LLM's context window into three segments to ensure 100\% recall of the specification.

\begin{figure*}[htbp]
\centerline{\includegraphics[width=1\linewidth]{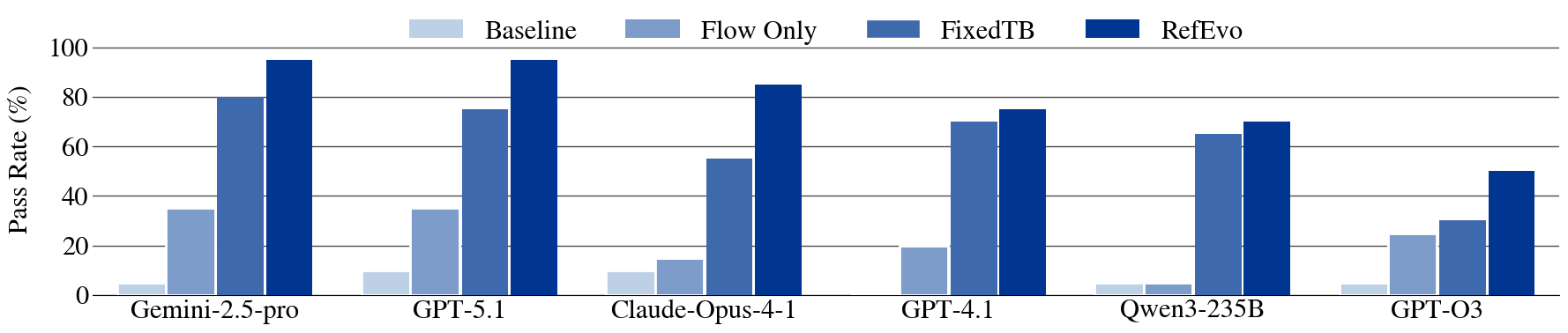}}
\caption{End-to-End Success Rate across different models and modes. RefEvo consistently outperforms baselines.}
\label{fig:main_results}
\end{figure*}

\begin{figure}[htbp]
\centerline{\includegraphics[width=1\linewidth]{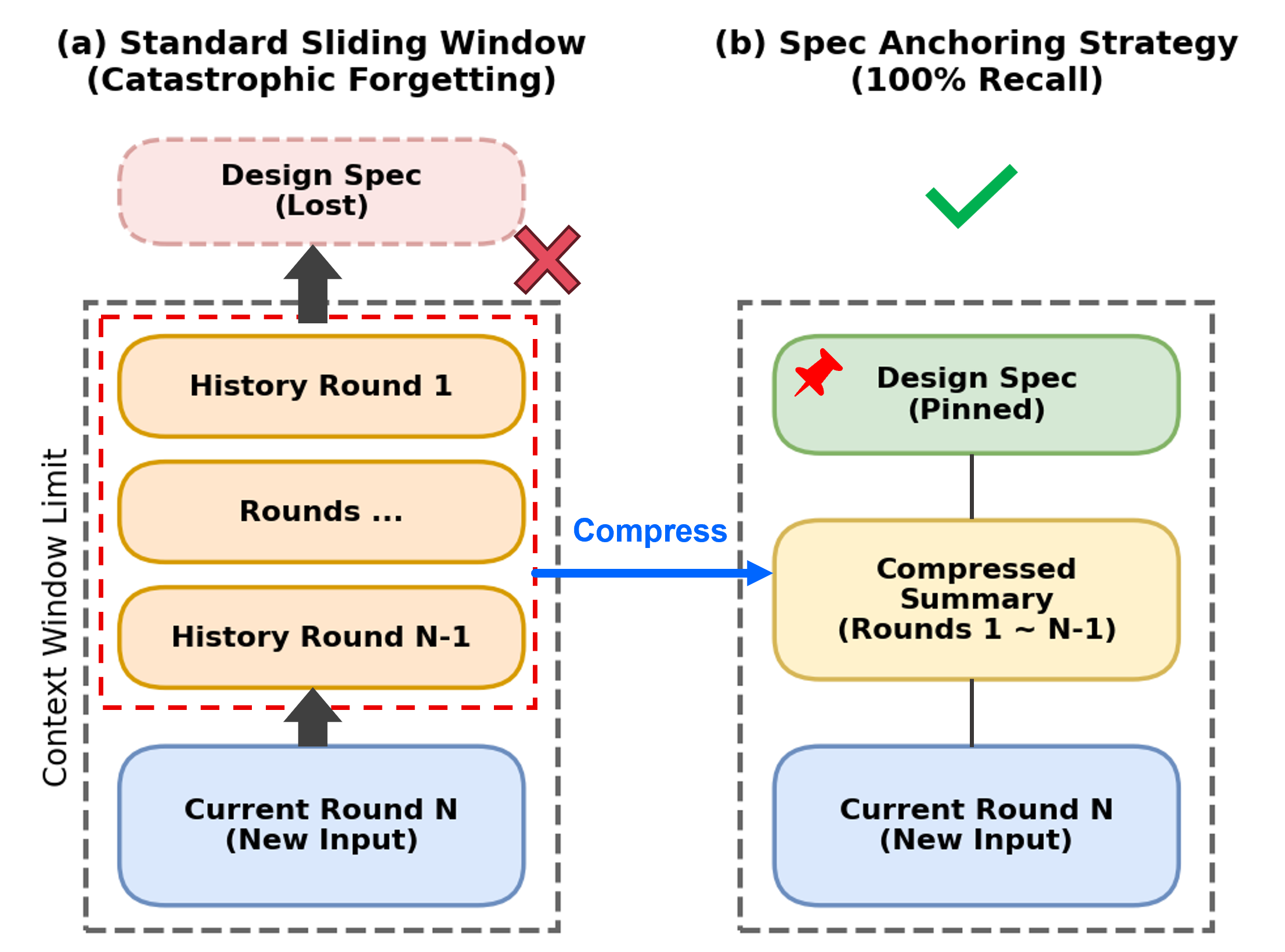}} 
\caption{Comparison of context management strategies. Spec Anchoring pins the specification as an immutable anchor and compresses historical logs, ensuring the Dialectical Arbiter always has access to the ground-truth oracle.}
\label{fig:spec_anchor}
\end{figure}

\begin{figure*}[htbp]
\centerline{\includegraphics[width=\linewidth]{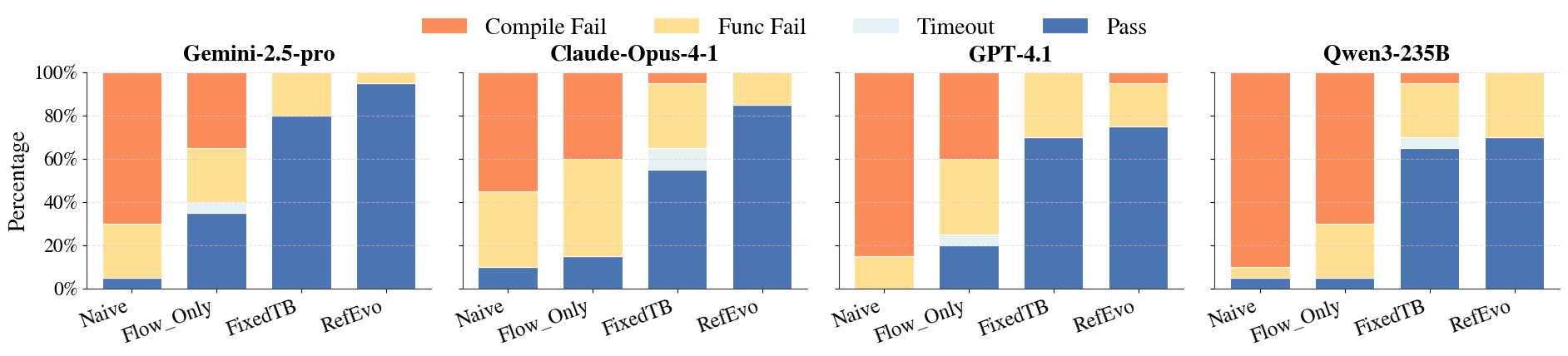}}
\caption{Failure distribution breakdown. The transition from Flow to FixedTB eliminates compilation errors, while the transition to RefEvo resolves functional mismatches via Co-Evolution.}
\label{fig:failure_dist}
\end{figure*}

\begin{itemize}
    \item \textbf{The Anchor (Immutable):} The initial Design Specification is pinned at the top of the context. It is never evicted during sliding window operations, serving as the constant \textbf{Oracle Reference} for Agent 4.
    \item \textbf{The Summary (Compressed):} Historical interaction rounds are summarized into concise state representations, preserving the ``reasoning chain'' while saving token budget.
    \item \textbf{The Workspace (Dynamic):} The most recent error logs and code snippets are placed here for active processing.
\end{itemize}
This partitioning allows the Dialectical Arbiter to maintain a consistent truth standard throughout the co-evolutionary process, even as the interaction history grows.

\section{Experiments and Results}
\label{sec:four}
\subsection{Experimental Setup}
We evaluated RefEvo on a benchmark of 20 hardware modules ranging from simple logic to complex protocols (e.g., AXI DMA, FFT, Keccak). The dataset was curated from open-source repositories including OpenCores\cite{opencores}, GenBen\cite{wan2025genbena}, and the XuanTie\cite{xuantie} RISC-V project. To ensure evaluation validity, all specifications were vetted by senior verification engineers. 

We tested five state-of-the-art LLMs, including Gemini-2.5-Pro, GPT-5.1, GPT-4.1, Qwen3, and Claude-Opus-4.1. We compared four experimental modes to isolate the contributions of our framework:
\begin{itemize}
    \item \textbf{Naive:} One-shot generation without any verification loop.
    \item \textbf{Flow\_Only:} Structured generation (Planning + Code Gen) without iterative refinement.
    \item \textbf{FixedTB:} iterative refinement enabled, but TB modification is strictly forbidden.
    \item \textbf{RefEvo (Ours):} The full framework with Co-Evolutionary Verification enabled.
\end{itemize}

For ground-truth validation, we employed a \textbf{SystemVerilog DPI-based co-simulation} framework. The generated SystemC models were compared against Golden RTL implementations under identical stimuli.

\subsection{Generation Capability Analysis}
Fig.~\ref{fig:main_results} illustrates the overall success rates. The \textbf{Naive} approach fails almost completely (0-10\%) due to pervasive syntax errors and incomplete code structures. Flow\_Only improves structural correctness, raising the pass rate to 15-35\%, yet functional bugs persist.
RefEvo achieves superior performance, with Gemini-2.5 and GPT-5.1 reaching a 95\% pass rate. Notably, for Claude-Opus-4.1, our mode outperforms Flow\_Only by 70\%. Complex modules like \textit{fpu\_div} and \textit{keccak} consistently fail in baselines but are successfully resolved in the RefEvo mode.

\begin{figure*}[htbp]
\centerline{\includegraphics[width=1\linewidth]{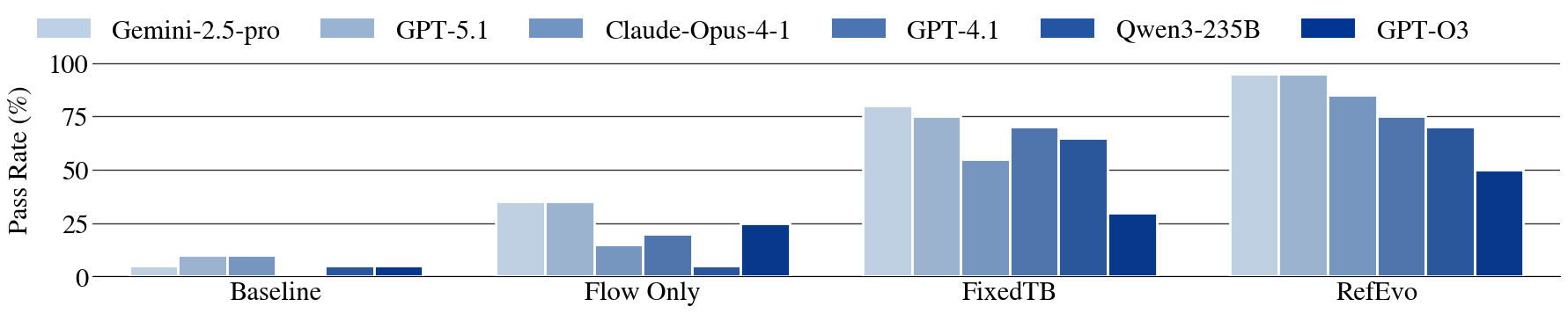}} 
\caption{Methodological Robustness Analysis. The consistent upward trend across all models confirms that RefEvo effectively enhances generation reliability independent of the underlying LLM's capability.}
\label{fig:robustness}
\end{figure*}

\begin{figure}[htbp]
\centerline{\includegraphics[width=1\linewidth]{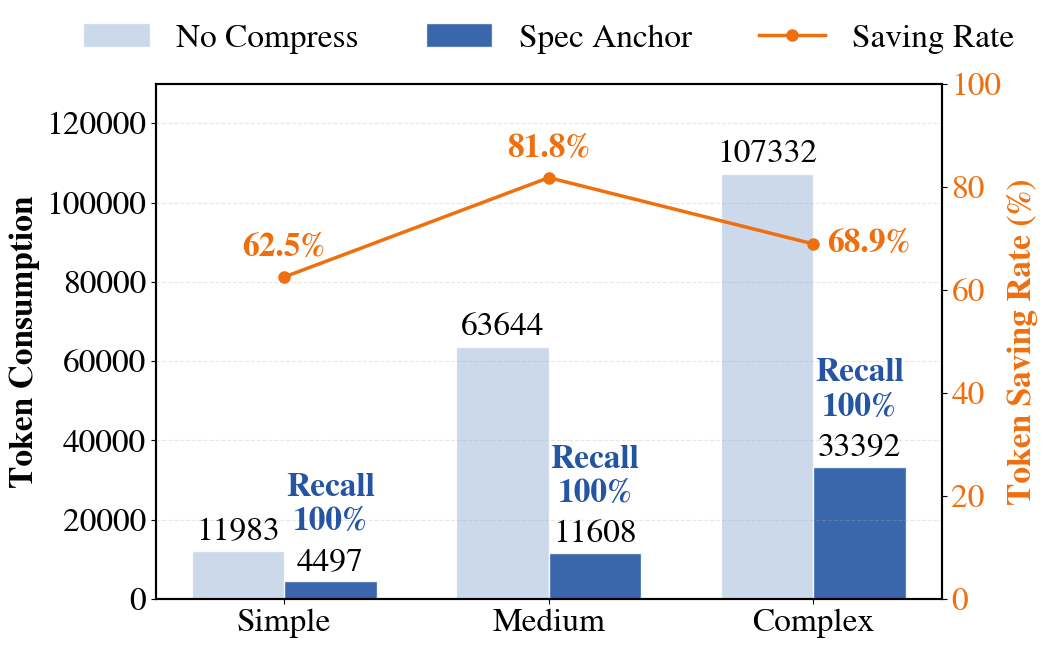}}
\caption{Token consumption comparison across Simple, Medium, and Complex design scales. RefEvo achieves an average reduction of \textbf{71.04\%}, with absolute savings scaling significantly for complex designs.}
\label{fig:token_eff}
\end{figure}

\subsection{Ablation Study: Mechanism Effectiveness}
To validate our verification strategy, we analyzed the failure distribution and the impact of Co-Evolutionary Verification (Fig.~\ref{fig:failure_dist}).

\begin{enumerate}
    \item \textbf{Impact of iterative refinement:} In Naive/Flow modes, the dominant failure mode is \textit{Compile Fail}. The introduction of the verification loop in FixedTB mode effectively eliminates these syntax errors, shifting the failure distribution towards \textit{Func Fail}.
    \item \textbf{Impact of Co-Evolution:} The comparison between FixedTB and RefEvo modes highlights the "Coupled Validation Failure" problem. In FixedTB mode, functional failures persist because the agent is forbidden from modifying the TB. When Co-Evolution is enabled, the \textit{Func Fail} rate drops significantly. This confirms that many failures were due to incorrect verification logic rather than flawed models, and our dialectical mechanism successfully resolved them.
\end{enumerate}

\subsection{Methodological Robustness}
To assess whether our framework's improvements are model-agnostic, we analyzed the performance trend across different LLMs. As shown in Fig.~\ref{fig:robustness}, despite the varying baseline capabilities of different models, the introduction of the RefEvo workflow consistently yields an upward trend in pass rates. This demonstrates that our method provides a robust enhancement component that effectively augments the capabilities of the underlying LLM.

\subsection{Efficiency Analysis: Context Compression}
To evaluate the scalability of our context management, we conducted controlled experiments across three design scales: \textit{Simple}, \textit{Medium}, and \textit{Complex}.

As shown in Fig.~\ref{fig:token_eff}, our Spec Anchoring strategy achieves an average token reduction of 71.04\% compared to the baseline. 
While the relative saving percentage peaks at 81.76\% for medium-scale designs, it remains substantial (68.89\%) for complex designs. The slight decrease in the savings ratio for the complex scenario is a deliberate architectural trade-off. To ensure 100\% Specification Recall---defined as the ratio of critical design constraints correctly retained across interaction turns---RefEvo intentionally preserves the full-length initial specification as an immutable anchor.
However, from the perspective of absolute efficiency, the benefit scales dramatically with complexity. RefEvo saves over 73,900 tokens in the complex scenario—nearly 10$\times$ more than in the simple case.

\section{Conclusion}
\label{sec:five}
This paper presented RefEvo, an agentic framework for agile reference model generation that bridges the gap between LLM capabilities and rigorous hardware verification standards. By integrating dynamic task planning with a co-evolutionary verification loop, RefEvo effectively addresses the challenges of semantic complexity and validation reliability. Our experiments demonstrate that the \textit{Dialectical Arbitration} mechanism significantly mitigates ``Coupled Validation Failure'', achieving a \textbf{95\%} success rate. Moreover, the Spec Anchoring strategy proves highly scalable, reducing token consumption by over \textbf{70\%}. RefEvo paves the way for fully automated, high-fidelity SoC verification, enabling a more efficient hardware-software co-design workflow.

\section*{Acknowledgments}
This work is supported by the National Natural Science Foundation of China under NSFC (Grant No. 92464301), the National Key Research and Development Program (Grant No. 2024YFB4405600), and the Key Research and Development Program of Jiangsu Province (Grant No. BG2024010).

\balance
\bibliographystyle{IEEEtran}
\bibliography{references}

\end{document}